\documentclass[12pt]{iopart}
\usepackage{graphicx}
\usepackage{setstack}
\usepackage{iopams}

\newcommand{\dfrac}[2]{\frac{\displaystyle #1}{\displaystyle #2}}
\newcommand{\email}[1]{\ead{#1}}
\newcommand{\affiliation}[1]{\address{#1}}

\newcommand{\vect}[1]{\boldsymbol{#1}}
\newcommand{\bm}[1]{\boldsymbol{#1}}
\newcommand{\umode}[2]{u_{#1,#2}}
\newcommand{\umodeflat}[2]{u^\flat_{#1,#2}}
\newcommand{\green}[1]{G_\xi\negthinspace\left(#1\right)}
\newcommand{\Dret}[2]{D_{#1}\negthinspace\left(#2\right)}
\newcommand{\Ylm}[2]{{f_{#1}^{\vect{#2}}}}
\newcommand{\besselJ}[1]{J{_{#1}}}
\newcommand{\heavy}[1]{\Theta\negthinspace\left(#1\right)}
\newcommand{\laplace}[2]{\mathcal{L}_{#1}\negthinspace\left\{#2\right\}}
\newcommand{\trace}[1]{\mathrm{Tr}\negthinspace\left(#1\right)}

\newcommand{\setZ}{\mathbb{Z}}

\newcommand{\ud}{\mathrm{d}}
\newcommand{\ue}{\mathrm{e}}
\newcommand{\uh}{\mathrm{h}}
\newcommand{\ub}{\mathrm{b}}

\newcommand{\um}{\mathrm{m}}
\newcommand{\ug}{\mathrm{g}}

\newcommand{\calC}{\mathcal{C}}
\newcommand{\calD}{\mathcal{D}}
\newcommand{\calF}{\mathcal{F}}
\newcommand{\calE}{\mathcal{E}}
\newcommand{\calH}{\mathcal{H}}
\newcommand{\calL}{\mathcal{L}}
\newcommand{\calP}{\mathcal{P}}
\newcommand{\calJ}{\mathcal{J}}
\newcommand{\calS}{\mathcal{S}}

\newcommand{\radim}{\rho}
\newcommand{\omadim}{\varpi}
\newcommand{\Uadim}{\bar{U}}
\newcommand{\Radim}{\bar{R}}
\newcommand{\Vadim}{\bar{V}}
\newcommand{\kappadim}{\bar{\kappa}}
\newcommand{\Lapbel}{\Delta_{\ncomp}}
\newcommand{\Boxadim}{\bar{\Box}_4}

\newcommand{\Mp}{M_{\mathrm{Pl}}}

\newcommand{\GeV}{\mathrm{GeV}}
\newcommand{\TeV}{\mathrm{TeV}}

\newcommand{\Ntot}{N}
\newcommand{\ncomp}{n}
\newcommand{\nextra}{n_{\ue}}
\newcommand{\Omegan}{\Omega_\ncomp}
\newcommand{\harm}{\Upsilon}

\newcommand{\vev}{v}
\newcommand{\higgs}{\phi}
\newcommand{\bhiggs}{\vect{\Phi}}
\newcommand{\gauge}{C}
\newcommand{\bgauge}{\vect{\calC}\negthinspace}

\newcommand{\bstrength}{\vect{\calH}}

\newcommand{\md}{m_\ud}
\newcommand{\mh}{m_\uh}
\newcommand{\mb}{m_\ub}
\newcommand{\mg}{m_\ug}
\newcommand{\gammag}{\gamma_\ug}

\newcommand{\eq}{\begin{equation}}
\newcommand{\qe}{\end{equation}}

\newcommand{\s}{\sigma}
\newcommand{\bo}{\omadim}

\begin{document}

\title{Graviton confinement inside hypermonopoles of any dimension}

\author{Se\'an Murray}
\email{sean.murray@uclouvain.be}
\affiliation{Institute of Mathematics and Physics, Centre for
Cosmology, Particle Physics and Phenomenology, Louvain University, 2 Chemin du
Cyclotron, 1348 Louvain-la-Neuve (Belgium)}

\author{Christophe Ringeval}
\email{christophe.ringeval@uclouvain.be}
\affiliation{Institute of Mathematics and Physics, Centre for
Cosmology, Particle Physics and Phenomenology, Louvain University, 2 Chemin du
Cyclotron, 1348 Louvain-la-Neuve (Belgium)}

\author{Simone Zonetti}
\email{simone.zonetti@uclouvain.be}
\affiliation{Institute of Mathematics and Physics, Centre for
Cosmology, Particle Physics and Phenomenology, Louvain University, 2 Chemin du
Cyclotron, 1348 Louvain-la-Neuve (Belgium)}

\date{\today}

\begin{abstract}
  We show the generic existence of metastable massive gravitons in the
  four-dimensional core of self-gravitating hypermonopoles in any
  number of infinite-volume extra-dimensions. Confinement is observed
  for Higgs and gauge bosons couplings of the order unity. Provided
  these resonances are light enough, they may realise the
  Dvali--Gabadadze--Porrati mechanism by inducing a four-dimensional
  gravity law on some intermediate length scales. The effective
  four-dimensional Planck mass is shown to be proportional to a
  negative power of the graviton mass. As a result, requiring gravity
  to be four-dimensional on cosmological length scales may solve the
  mass hierarchy problem.
\end{abstract}
\pacs{04.50.+h, 11.10.Kk, 98.80.Cq}
\maketitle

\section{Introduction}

The old and more recent interest in the existence of space-like
extra-dimensions has led to three main ways of accommodating their
presence with an apparent four-dimensional world~\cite{Nordstrom:1988fi,
  Kaluza:1921, Klein:1926tv}. In String Theory, extra-dimensions are
compactified and can be probed only at an energy scale exceeding their
inverse radius. The non-observation of the associated Kaluza--Klein
(KK) excitations at colliders have pushed such a scale above the
$\TeV$. However, motivated by orbifold constructions, it has been
realised that the typical size of the extra-dimensions probed by
gravity could be much larger than the one felt by the gauge (and
matter) fields~\cite{Antoniadis:1990ew, Horava:1996qa,
  Lukas:1998qs}. In this picture, our universe could be a
four-dimensional brane on which the Standard Model particles are
confined, embedded in a higher-dimensional bulk
~\cite{Arkani-Hamed:1998rs, Antoniadis:1998ig,
  Randall:1999ee,Gregory:2000jc, Ringeval:2001cq}. Still, gravity is
tested to be four-dimensional on length-scales ranging from the
micrometers to the cosmological distances~\cite{Fischbach:2001ry,
  Bezerra:2010pq}. A micrometer is nevertheless much larger than a
$\TeV^{-1}$ while the current cosmic acceleration might be the
signature that gravity is actually no longer standard on the largest
length scales~\cite{Deffayet:2001pu}. These motivations have led to
two other gravity confinement mechanisms. In the Randall--Sundrum (RS)
models, the extra-dimensions felt by gravity are non-compact but still
of finite volume such that deviations from four-dimensional physics
appear only at lengths smaller than their curvature
radius~\cite{Randall:1999vf}. In the Dvali--Gabadadze--Porrati (DGP)
models, the extra-dimensions are non-compact and of infinite
volume. Gravity can be observed as four-dimensional provided the
graviton flux between two masses is prevented to leak into the
extra-dimensions. In the original DGP model, this is obtained by
adding a (quantum induced) four-dimensional Einstein--Hilbert term on
the brane~\cite{Dvali:2000hr,Dvali:2000xg}. Although it has been
argued that some instabilities should show up in the original
approach~\cite{Deffayet:2006wp, Gregory:2007xy}, stable extensions
have been proposed since~\cite{Gabadadze:2003ck, Kaloper:2007ap,
  Kaloper:2007qh, Dvali:2007kt, deRham:2007rw, deRham:2007xp,
  deRham:2009wb, deRham:2010rw}. The so-called regularised DGP models
assume an ad-hoc varying gravitational coupling constant along the
extra-dimensions such that gravitons are reflected back onto the brane
in the same way as a varying dielectric constant reflects
photons. Assuming the gravity action to be
\begin{equation}
\label{eq:action_reg}
S= \dfrac{1}{2\kappa^2} \int{
  \sqrt{|g|} \,R \, \calF(X) \, \ud^{\nextra+4} X},
\end{equation}
where $\nextra$ stands for the number of extra-dimension, $R$ is the
Ricci scalar and $g$ is the determinant of the metric tensor, it has
been shown that $4D$-like gravity can be recovered provided the
function $\calF(X)$ is peaked enough in a narrow region around the
brane~\cite{Kolanovic:2003am, Kolanovic:2003da}. However, the
existence of a similar mechanism in a well-defined physical framework
is a non-trivial issue. Indeed, the role of the function $\calF$ in
Eq.~(\ref{eq:action_reg}) could be played by $g(X)$ if there are
non-vanishing stress tensor sources in the
bulk~\cite{Shaposhnikov:2004ds}. A varying Planck mass can also be
obtained by considering a dilaton $\psi$ that would condense on the
brane such that $\calF(X) = \exp[\psi(X)]$ has the required
profile. In any case, none of these fields would be independent and
their equations of motion have to be solved in a given field
theoretical setup to address this question.

This has been done in Refs.~\cite{Ringeval:2004ju, DeFelice:2008af} in
a scalar-tensor theory of gravity by considering our brane to be a
self-gravitating topological defect~\cite{Akama:1982jy,
  Rubakov:1983bb, Visser:1985qm, Gibbons:1986wg, Cvetic:1996vr}. The
core is a flat four-dimensional spacetime supposed to be our universe
while the non-trivial defect-forming field configurations curve the
extra-dimensions. Such a defect can be formed by the spontaneous
breakdown of a local symmetry in the bulk whose stress-tensor ends up
being non-vanishing only in a region defining the brane
thickness. Typically, it is given by the Compton wavelength of the
various defect-forming fields. The extra-dimensional spacetime is
asymptotically flat and of infinite volume, as in the DGP model. In
six dimensions, this can be obtained by breaking a local $U(1)$
symmetry to form a hyperstring whose core has three spatial dimensions
(there is $\nextra=2$ extra-dimensions). The seven-dimensional version
is a 't~Hooft-Polyakov hypermonopole obtained by breaking an $SO(3)$
symmetry in $\nextra=3$ extra-dimensions. As shown in
Ref.~\cite{Ringeval:2004ju}, reflecting gravitons onto the defect core
requires a violation of the positivity energy conditions in General
Relativity~\cite{Wald:1984rg}. This happens to be impossible for the
hyperstring unless one adds a source of negative energy in the
bulk. As a matter of fact, a negative cosmological constant does the
job and the model becomes of the RS type, with finite volume
extra-dimensions and a severe fine-tuning problem~\cite{Roessl:2002rv,
  Peter:2003zg}. It is well known that a constant positive curvature
term in the Einstein equations behaves like a perfect fluid with a
negative equation of state parameter and can therefore mimic matter
with negative
pressure~\cite{Carroll:2001ih}. Ref.~\cite{DeFelice:2008af} has shown
that this mechanism is indeed at work inside the seven-dimensional
hypermonopole: assuming isotropy, the positive curvature of the
two-dimensional orthoradial extra-dimensions acts as a potential
barrier for the propagation of gravitons. These become resonant and
massive. The next question is therefore to determine if curving more
than two orthoradial extra-dimensions still allows for a similar
gravity confinement in higher dimensional spacetimes.

The present article is devoted to this issue. In the following, we
show that the DGP-like graviton confinement mechanism by curvature
effects is indeed a generic feature of self-gravitating hypermonopoles
and occurs in any asymptotically flat spacetime of strictly more than
six dimensions. More precisely, in $4+\nextra$ dimensions,
hypermonopoles can be formed by the spontaneous breakdown of an
$SO(\nextra)$ symmetry to $SO(\nextra-1)$. Moreover, in order to allow
for a varying Planck mass in the bulk, gravity is supposed to be of
the scalar-tensor type, $\psi$ being the dilaton. After having derived
and numerically solved the equations of motion in Sec.~\ref{sec:bg},
we show that the extra-dimensions are of infinite volume and
asymptotically flat while being strongly positively curved in an
intermediate region. Such a configuration is obtained without
fine-tuning and naturally occurs for values of the field coupling
constant of the order unity. In Sec.~\ref{sec:spintwo}, we solve for
the propagation of spin-two fluctuations along the brane and find
strongly peaked massive metastable resonances. We then illustrate how
these resonances may realise the DGP mechanism by deriving the resulting
Newtonian potential on the brane: it remains $(4+\nextra)$-dimensional
at small and large distances but behaves as $d$-dimensional in an
intermediate range with $d<4+\nextra$. Finally, we discuss the mass
hierarchy problem and show that the four-dimensional effective Newton
constant is proportional to a positive power of the mass $\mg$ of the
lightest associated graviton resonances. This property is analogous to
the existence of a cross-over distance at which gravity becomes
$\Ntot$-dimensional in the DGP regularised
models~\cite{Kolanovic:2003am, Kolanovic:2003da}. In terms of the
Planck mass, we find in Eq.~(\ref{eq:Max}) that
\begin{equation}
\Mp^2 \propto \dfrac{M^{\nextra+2}}{\mg^{\nextra}}\,,
\end{equation}
where $M$ stands for the $\Ntot$-dimensional Planck mass. Since
four-dimensional gravity requires extremely light resonances, our
mechanism necessarily implies a very small four-dimensional gravity
coupling constant. Finally, concerning the smallest length scales, we
show that the distance under which gravity is again
$\Ntot$-dimensional depends on the gravitational redshift induced by
the hypermonopole forming fields. It can be made arbitrarily small,
independently of the graviton masses, provided the Higgs and gauge
bosons have masses close to the Planck mass $M$ in $\Ntot$ dimensions.

\section{Hypermonopoles of any dimension}
\label{sec:bg}

In this section, we assume our universe to be the four-dimensional
core of a hypermonopole living in $\Ntot=4+\nextra$ dimensions. This
topological defect can be formed by spontaneously breaking an
$SO(\nextra)$ symmetry to $SO(\nextra-1)$. We impose this symmetry to
be local such that the defect does not exhibit long range interactions
and has a localised stress tensor allowing asymptotically flat
extra-dimensions. As mentioned in the introduction, gravity is assumed
to be of the scalar-tensor type to allow for a varying effective
Planck mass along the extra-dimensions. In the Jordan frame, the
action describing this system reads
\begin{eqnarray}
   S  &=&  \dfrac{1}{2 \kappa^2} \int e^\psi \sqrt{|g|} \left[R
    -g^{AB} \partial_A \psi \partial_B \psi -U(\psi) \right] \ud^\Ntot
  x  \nonumber \\ & & +  \int \sqrt{|g|} \left[- \dfrac{1}{2}g^{AB} \calD_A \bhiggs \cdot
    \calD_B \bhiggs  +  \dfrac{1}{8} \trace{\bstrength_{AB}
      \bstrength^{AB}} \right. \nonumber \\ & & \left. - \dfrac{\lambda}{8}\left( \bhiggs \cdot \bhiggs - \vev^2
    \right)^2\right] \ud^\Ntot x,
\label{eq:action}
\end{eqnarray}
where the $SO(\nextra)$ Higgs field is an $\nextra$-dimensional vector
$\bhiggs=\{\higgs^a\}$ with $a\in\{1,2,\dots,\nextra\}$. Gauge
invariance under local $SO(\nextra)$ transformations is ensured by
defining the covariant derivatives as
\begin{equation}
\label{eq:covD}
  \calD_A \bhiggs \equiv \partial_A \bhiggs - q \bgauge_A \bhiggs,
\end{equation}
where $q$ is the Higgs charge and
$\bgauge_A=\{(\gauge_A)^{ab}\}$ are the gauge field
matrices. The associated field strength tensor matrices are given by
\begin{equation}
  \bstrength_{AB} \equiv \partial_A \bgauge_B - \partial_B \bgauge_A -
  q \left[\bgauge_A, \bgauge_B \right].
\end{equation}
The last term in Eq.~(\ref{eq:action}) is the Higgs potential
$V(\bhiggs)$ and breaks $SO(\nextra)$ to $SO(\nextra-1)$ such that the
topology of the vacuum manifold is the same as the $\ncomp$-sphere
$S^{\ncomp} \cong SO(\nextra)/SO(\nextra-1)$, where we have defined
\begin{equation}
\ncomp \equiv \nextra-1.
\end{equation}
Since the $\ncomp^{\mathrm{th}}$ homotopy group
$\pi_\ncomp(S^\ncomp)=\setZ$ is non-trivial, we expect the formation
of hypermonopoles mapping $SO(\nextra)$ to the $\nextra$
extra-dimensions~\cite{Kibble:1976}. In Eq.~(\ref{eq:action}), the
quantity $U(\psi)$ encodes the dilaton potential in the Jordan
frame. For simplicity, we choose the dilaton to be a free massive
particle of mass $\md$ in the Einstein frame\footnote{in which the
  scalar and tensor degrees of freedom are decoupled} such that
\begin{equation}
U(\psi) = \md^2 e^{2 \psi/(\ncomp+3)}\,.
\end{equation}
One can check that the above equations reduce to the hyperstring of
Ref.~\cite{Ringeval:2004ju} for $\nextra=2$ and to the
't~Hooft--Polyakov hypermonopole of Ref.~\cite{DeFelice:2008af} when
$\nextra=3$.

In the following, after having introduced our Ansatz for the field
profiles, we derive and solve the equations of motion assuming
isotropic extra-dimensions.

\subsection{Equations of motion}

\subsubsection{Gravity sector}

Varying the action with respect to the metric tensor and the dilaton
gives the Einstein--Jordan equations
\begin{eqnarray}
\label{eq:tensor}
G_{AB} &=& \ue^{-\psi} \kappa^2 T_{AB} + 2 \partial_{A}
\psi \partial_{B} \psi - \dfrac{1}{2} g_{AB} \left[ 3
  \partial_{X} \psi \partial^{X} \psi \right. \nonumber \\ & & \left. +
  U(\psi) \right] + \nabla_{A} \partial_{B} \psi -
g_{AB} \Box \psi, \\
\label{eq:scalar}
\Box \psi & =& \dfrac{1}{2} U(\psi) + \dfrac{1}{2}
\dfrac{\ud U}{\ud \psi} - \dfrac{1}{2} R - \dfrac{1}{2} \partial_{X} \phi
\partial^{X} \psi,
\end{eqnarray}
where $\Box \equiv \nabla_{X}
\partial^{X}$, $G_{AB}$ is the $\Ntot$-dimensional Einstein tensor and
$T_{AB}$ the matter stress tensor
\begin{equation}
\label{eq:tmunugen}
T_{AB} = \calD_A \bhiggs \cdot \calD_B \bhiggs - \dfrac{1}{2} g^{CD}
\trace{\bstrength_{AC}\bstrength_{BD}} + g_{AB} \calL_\um,
\end{equation}
where $\calL_\um$ is the Higgs and gauge field Lagrangian.

\subsubsection{Matter sector}

The variations of Eq.~(\ref{eq:action}) with respect to the Higgs and
gauge fields gives the Klein--Gordon and Maxwell-like equations:
\begin{equation}
\label{eq:higgs}
  \Box \bhiggs - q\, g^{AB} \left[ \left(\nabla_A \bgauge_B
    \bhiggs\right) + \bgauge_B \calD_A \bhiggs \right]-
  \dfrac{\partial V}{\partial \bhiggs} = 0,
\end{equation}
and
\begin{eqnarray}
\label{eq:gauge}
  \left( \nabla_A \bstrength^{AB}\right)^{ab} & + & q\, g^{AB}
  \left[\left(\calD_A \bhiggs\right)^a \higgs^b - \left(\calD_A
      \bhiggs \right)^b
    \higgs^a \right] \\
  & + & q \left(\bstrength^{AB}\cdot \bgauge_A - \bgauge_A \cdot
    \bstrength^{AB}\right)^{ab} = 0.
\end{eqnarray}

\subsubsection{Metric and field Ansatz}

Respecting the hyperspherical static symmetry in the $\nextra$
extra-dimensions and Poincar\'e invariance along the brane gives the
metric
\begin{equation}
\label{eq:metric}
  \ud s^2 = g_{AB} \ud x^A \ud x^B = e^{\sigma(r)} \eta_{\mu\nu} \ud
  x^\mu \ud x^\nu + \ud r^2 + \omega^2(r) \ud \Omegan^2.
\end{equation}
The metric element over $S^\ncomp$ being
\begin{equation}
\ud \Omegan^2 =  \sum_{i=1}^n \harm_i(\theta_{j<i}) \ud \theta_i^2,
\end{equation}
with $\harm_1=1$, $\harm_2=\sin^2(\theta_1)$ and
\begin{equation}
\harm_i(\theta_{j<i}) \equiv \harm_2(\theta_1)\prod_{j=2}^{i-1}
\sin^2(\theta_j)\quad \mathrm{for~~}i=3,\ldots,n~.
\end{equation}
For a defect configuration, the Higgs field vanishes in the core
whereas it asymptotically recovers its vacuum expectation
value. Enforcing the spacetime symmetries, we assume a radial field
such that
\begin{equation}
\label{eq:anshiggs}
  \higgs^a = \varphi(r) \dfrac{x^a}{r}\,,
\end{equation}
with $\varphi(0)=0$ and $\varphi(r)\rightarrow \vev$ when $ r
\rightarrow \infty$. In Eq.~(\ref{eq:anshiggs}), $r^2 = \delta_{ab}
x^a x^b$ where the $\{x^a\}$ stands for the $\nextra$ Cartesian
coordinates defined by
\begin{eqnarray}
  x^1 &=& r \cos\theta_1, \quad x^2 = r \sin \theta_1 \cos \theta_2, \\
  &\vdots& \nonumber \\
  x^{\ncomp} &=& r \sin\theta_1\sin\theta_2 \dots \sin\theta_{\ncomp-1}
  \cos \theta_\ncomp, \nonumber \\
  x^{\nextra} &=& r \sin\theta_1\sin\theta_2 \dots
  \sin\theta_{\ncomp-1} \sin \theta_\ncomp.
\end{eqnarray}
We also assume that the dilaton depends only on the radial coordinate
\eq \psi=\psi(r)~.\label{dilaton} \qe
Our Ansatz for the gauge field is the generalisation of the
't~Hooft--Polyakov configuration~\cite{'tHooft:1974qc,Polyakov:74},
with a unity winding number. Requiring the stress tensor to vanish at
infinity imposes the covariant derivative to vanish. {}From
Eqs.~(\ref{eq:covD}) and (\ref{eq:anshiggs}), one gets
\begin{equation}
\label{eq:ansgauge}
(\gauge_{\theta_i})^{ab} = \dfrac{1 - Q(r)}{q r^2} \left(
  \dfrac{\partial x^a} {\partial \theta_i} x^b - \dfrac{\partial
    x^b}{\partial \theta_i} x^a \right),
\end{equation}
where the dimensionless function $Q(0)=1$ for regularity in the core
and $Q(r) \rightarrow 0$ at infinity. All the other $\bgauge_A$ are
vanishing. Under such an Ansatz, observing that
\begin{equation}
  \trace{\bgauge_{\theta_i} \bgauge_{\theta_j}} = -2 \left[
    \dfrac{1-Q(r)}{q \omega(r)}\right]^2 g_{\theta_i \theta_j},
\end{equation}
the gravity and matter sector equations considerably simplify and we
write down only the final result in the next section (see the appendix
for some intermediate steps).

\subsubsection{Dimensionless equations}

For convenience, we introduce the following dimensionless
quantities. The radial distance can be expressed in unit of the Higgs
Compton wavelength such that
\begin{equation}
\radim \equiv \mh r = \sqrt{\lambda} \vev \,r,
\end{equation}
where $\mh$ is the mass of the Higgs boson. Similarly, the
dimensionless angular metric coefficient and Higgs field are defined
by
\begin{equation}
  \omadim\equiv \mh \omega, \qquad f \equiv \dfrac{\varphi}{v}\,.\label{eq:fdefinition}
\end{equation}
The gravity, Higgs and gauge coupling constants account for three
dimensionless parameters in the equations of motion (\ref{eq:tensor})
to (\ref{eq:gauge}) which can be recast as
\begin{equation}
\alpha \equiv \kappa^2 \vev^2, \qquad \epsilon\equiv \dfrac{q^2
  \vev^2}{\lambda \vev^2} = \dfrac{\mb^2}{\mh^2}, \qquad \beta \equiv
\dfrac{\md^2}{\lambda \vev^2}=\dfrac{\md^2}{\mh^2}\,,
\end{equation}
$\mb$ being the mass of the gauge bosons. After some rather long
algebra, a dot denoting differentiation with respect to $\radim$, the
dimensionless equations of motion in the gravity sector read
\begin{eqnarray}
\label{eq:munu}
\dfrac{3}{2} \ddot{\sigma} & + & \dfrac{3}{2} \dot{\sigma}^2 + \ncomp
\dfrac{\ddot{\omadim}}{\omadim} + \ncomp(\ncomp-1)
\dfrac{\dot{\omadim}^2 - 1}{2\omadim^2} + \dfrac{3}{2} \ncomp
\dot{\sigma} \dfrac{\dot{\omadim}}{\omadim} \nonumber \\
& = & -\alpha e^{\-\psi} \calE - \ddot{\psi} - \dfrac{3}{2} \dot{\psi}^2
- \left(\dfrac{3}{2} \dot{\sigma} + \ncomp
  \dfrac{\dot{\omadim}}{\omadim}\right)
\dot{\psi}  - \dfrac{\Uadim}{2}\,,\\
\label{eq:rr}
\dfrac{3}{2} \dot{\sigma}^2 & + & \ncomp(\ncomp-1)
\dfrac{\dot{\omadim}^2 - 1}{2\omadim^2} + 2 \ncomp \dot{\sigma}
\dfrac{\dot{\omadim}}{\omadim} \nonumber \\ & = & \alpha e^{-\psi}
\calP + \dfrac{1}{2} \dot{\psi}^2 - \left(2 \dot{\sigma} + \ncomp
  \dfrac{\dot{\omadim}}{\omadim} \right) \dot{\psi} -
\dfrac{\Uadim}{2}\,,\\
\label{eq:theta}
2 \ddot{\sigma} & + & \dfrac{5}{2} \dot{\sigma}^2 + (\ncomp-1)
\dfrac{\ddot{\omadim}}{\omadim} + (\ncomp-1)(\ncomp-2)
\dfrac{\dot{\omadim}^2 -1}{2\omadim^2}  +  2 (\ncomp-1)
\dot{\sigma} \dfrac{\dot{\omadim}}{\omadim}\nonumber \\ & = & - \alpha e^{-\psi}
\calE_\perp - \ddot{\psi} - \dfrac{3}{2} \dot{\psi}^2 - 
\left[2 \dot{\sigma} + (\ncomp-1)\dfrac{\dot{\omadim}}{\omadim}
\right] \dot{\psi} - \dfrac{\Uadim}{2}\,,\\
\label{eq:dilaton}
\ddot{\psi} & + & \dfrac{1}{2} \dot{\psi}^2 + \left(2 \dot{\sigma} +
  \ncomp \dfrac{\dot{\omadim}}{\omadim} \right) \dot{\psi} =
\dfrac{1}{2} \left(\Uadim + \dfrac{\ud\Uadim}{\ud \psi} - \Radim \right).
\end{eqnarray}
The dimensionless dilaton potential is $\Uadim = U/\mh^2$ and the
dimensionless Ricci scalar $\Radim = R/\mh^2$ stands for
\begin{equation}
\label{eq:ricci}
  \Radim = -4 \ddot{\sigma} - 5 \dot{\sigma}^2 - 2 \ncomp
  \dfrac{\ddot{\omadim}}{\omadim} - \ncomp(\ncomp-1)
  \dfrac{\dot{\omadim}^2 - 1}{\omadim^2} - 4 \ncomp \dot{\sigma}
  \dfrac{\dot{\omadim}}{\omadim}\,.
\end{equation}
The quantities $\calE$ and $\calP$ are respectively the energy density
and pressure generated by the Higgs and gauge fields along the radial
extra-dimension~\footnote{We also have $\calE=-\calL_\um/(\mh^2
  \vev^2)$.}
\begin{eqnarray}
  \calE  & = & \dfrac{\dot{f}^2}{2} + \dfrac{\ncomp\dot{Q}^2 }{2\epsilon
    \omadim^2} + \dfrac{\ncomp}{2\omadim^2} f^2 Q^2 +
  \dfrac{\ncomp(\ncomp-1)}{4
    \epsilon \omadim^4} (1-Q^2)^2 + \Vadim , \\
  \calP  & = & \dfrac{\dot{f}^2}{2} + \dfrac{\ncomp \dot{Q}^2}{2\epsilon
    \omadim^2} - \dfrac{\ncomp}{2\omadim^2} f^2 Q^2 -
  \dfrac{\ncomp(\ncomp-1)}{4 \epsilon \omadim^4} (1-Q^2)^2- \Vadim ,
\end{eqnarray}
where $\Vadim(f)$ stands for the dimensionless Higgs potential
\begin{equation}
\Vadim(f) \equiv \dfrac{1}{8} \left(1- f^2\right)^2.
\end{equation}
The quantity $\calE_\perp$ appearing in Eq.~(\ref{eq:theta}) is the
energy density along the orthoradial directions and reads
\begin{equation}
\label{eq:enerperp}
  \calE_\perp = \dfrac{\dot{f}^2}{2}  + \dfrac{(\ncomp-2)\dot{Q}^2
  }{2\epsilon \omadim^2} + \dfrac{\ncomp-2}{2\omadim^2} f^2 Q^2  +
  \dfrac{(\ncomp-1)(\ncomp-4)}{4 \epsilon \omadim^4} (1-Q^2)^2+ \Vadim.
\end{equation}
The dynamical equations for $f$ and $Q$ stem from
Eqs.~(\ref{eq:higgs}) and (\ref{eq:gauge})
\begin{eqnarray}
\label{eq:f}
  \ddot{f} & + & \left(2 \dot{\sigma} + \ncomp
    \dfrac{\dot{\omadim}}{\omadim} \right) \dot{f} - \ncomp \dfrac{f
    Q^2}{\omadim^2} - \dfrac{\ud \Vadim}{\ud f} = 0, \\
\label{eq:Q}
  \ddot{Q} & + & \left[2 \dot{\sigma} + (\ncomp-2)
    \dfrac{\dot{\omadim}}{\omadim} \right] \dot{Q} +
  \dfrac{\ncomp-1}{\omadim^2} (Q-Q^3) - \epsilon f^2 Q = 0.
\end{eqnarray}
These equations match with those of the six-dimensional hyperstring
and seven-dimensional hypermonopole derived in
Refs.~\cite{Ringeval:2004ju, DeFelice:2008af}. Let us however notice
the presence of new terms for $\ncomp>2$ in the orthoradial
equation~(\ref{eq:theta}) as well as in the associated stress energy in
Eq.~(\ref{eq:enerperp}). It is also worth remarking from
Eq.~(\ref{eq:enerperp}) that for $\ncomp=2$ or $3$, the gauge field is
generating a negative potential in the orthoradial extra-dimensions,
which becomes again positive for $\ncomp > 4$.

\subsection{Background fields and geometry}

\subsubsection{Boundary conditions}

The boundary conditions for the metric coefficients and fields are
fixed by requiring regularity in the core and a Dirac hypermonopole
configuration asymptotically. As already discussed, we look for
asymptotically flat spacetime, i.e.
\begin{eqnarray}
\label{eq:boundasymp}
  \lim_{\radim \rightarrow \infty} f(\radim) & = 1, \qquad
  \lim_{\radim \rightarrow
    \infty} Q(\radim) =0, \nonumber \\
  \lim_{\radim \rightarrow \infty} \sigma(\radim) & = 0, \qquad
  \lim_{\radim \rightarrow \infty} \dfrac{\omadim(\radim)}{\radim}=
  1, \qquad \lim_{\radim \rightarrow \infty} \psi(\radim) = 0.
\end{eqnarray}
Notice that $\sigma$ could be shifted by a constant value since all
the equations of motion depend only on $\dot{\sigma}$: this reflects
the expected invariance with respect to a rescaling of the internal
brane coordinates $x^\mu$. We have also chosen the dilaton to vanish
at infinity since this minimises its potential energy and $\psi=0$ is
an exact solution of Eq.~(\ref{eq:dilaton}) for $\Radim=0$. One can
check that this last condition is indeed asymptotically fulfilled with
the limits of Eq.~(\ref{eq:boundasymp}). Let us notice that the metric
far from the core is not a generalisation of the conical flat metric
existing around a cosmic string. As can be checked in
Eq.~(\ref{eq:ricci}), as soon as $\ncomp(\ncomp-1)\neq 0$ one has
$\omadim=\radim$, i.e. there is no missing angle (for
$\dot{\sigma}=0$).

In the hypermonopole core, the $SO(\nextra)$ symmetry should be
restored and the spacetime geometry has to be regular. As a result,
the fields satisfy
\begin{eqnarray}
\label{eq:boundcore}
\lim_{\radim \rightarrow 0} f(\radim) & = 0, \qquad \lim_{\radim \rightarrow
    0} Q(\radim) = 1, \nonumber \\
 \lim_{\radim \rightarrow 0} \dot{\sigma}(\radim) & = 0, \qquad
  \lim_{\radim \rightarrow 0} \dfrac{\omadim(\radim)}{\radim} =
  1, \qquad \lim_{\radim \rightarrow 0} \dot{\psi}(\radim) = 0.
\end{eqnarray}

\subsubsection{Solutions}

\begin{figure}
\begin{center}
\includegraphics[width=0.7\textwidth]{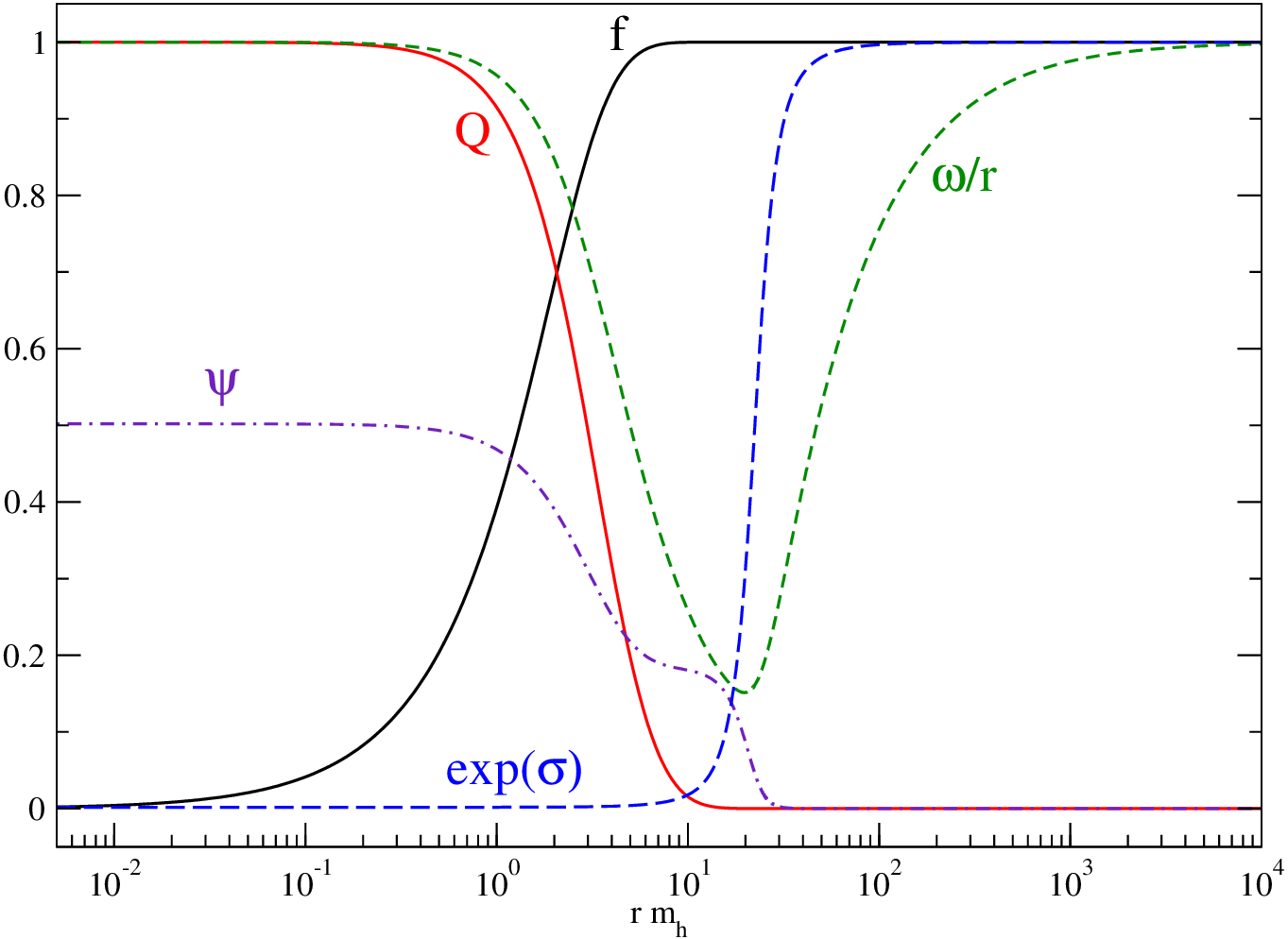}
\caption{Field and metric profiles forming an eight-dimensional
  ($\ncomp=3$) hypermonopole for $\alpha=2.60$, $\epsilon=0.25$ and
  $\beta=1.00$ (top). The space-time is flat in the core and at
  infinity but strongly curved in the intermediate region. The
  dilaton condenses in the core and traces the energy density
  distribution. See also Fig.~\ref{fig:bgn3metric}.}
\label{fig:bgn3}
\end{center}
\end{figure}

{}From Eqs.~(\ref{eq:boundasymp}) and (\ref{eq:boundcore}), we have
ten boundary conditions to solve the ten-dimensional first order
non-linear differential system that can be obtained from the second
order Eqs.~(\ref{eq:munu}), (\ref{eq:theta}), (\ref{eq:dilaton}),
(\ref{eq:f}) and (\ref{eq:Q}). Notice that Eq.~(\ref{eq:rr}) is not
included since this is a constraint equation and is redundant with the
previous set, up to a constant which is fixed once the boundary
conditions are specified. Finding numerical solutions of this system
is non-trivial and we have used the conditioning mesh methods
implemented in Ref.~\cite{Cash:2005aa}. We have first checked our
numerical implementation by recovering the $\ncomp=1$ and $\ncomp=2$
solutions of Refs.~\cite{Ringeval:2004ju, DeFelice:2008af} before
solving the system for $\ncomp>2$. Despite the new terms appearing in
the equations of motion, we have found hypermonopole solutions for any
tested value of $\ncomp$. All of them exhibit similar patterns than
those found in seven-dimensions. For coupling constant of order unity,
the spacetime is strongly curved in an intermediate region where the
field derivatives are non-vanishing, and in particular $\omadim$
remains almost stationary with respect to $\radim$. As a result, the
hypersurface of the $\ncomp$-sphere of radius $\radim$ becomes
constant and the extra-dimensions are cylindrically shaped. At further
distances, $\omadim \sim \radim$ again and the spacetime becomes flat.

\begin{figure}
\begin{center}
\includegraphics[width=0.7\textwidth]{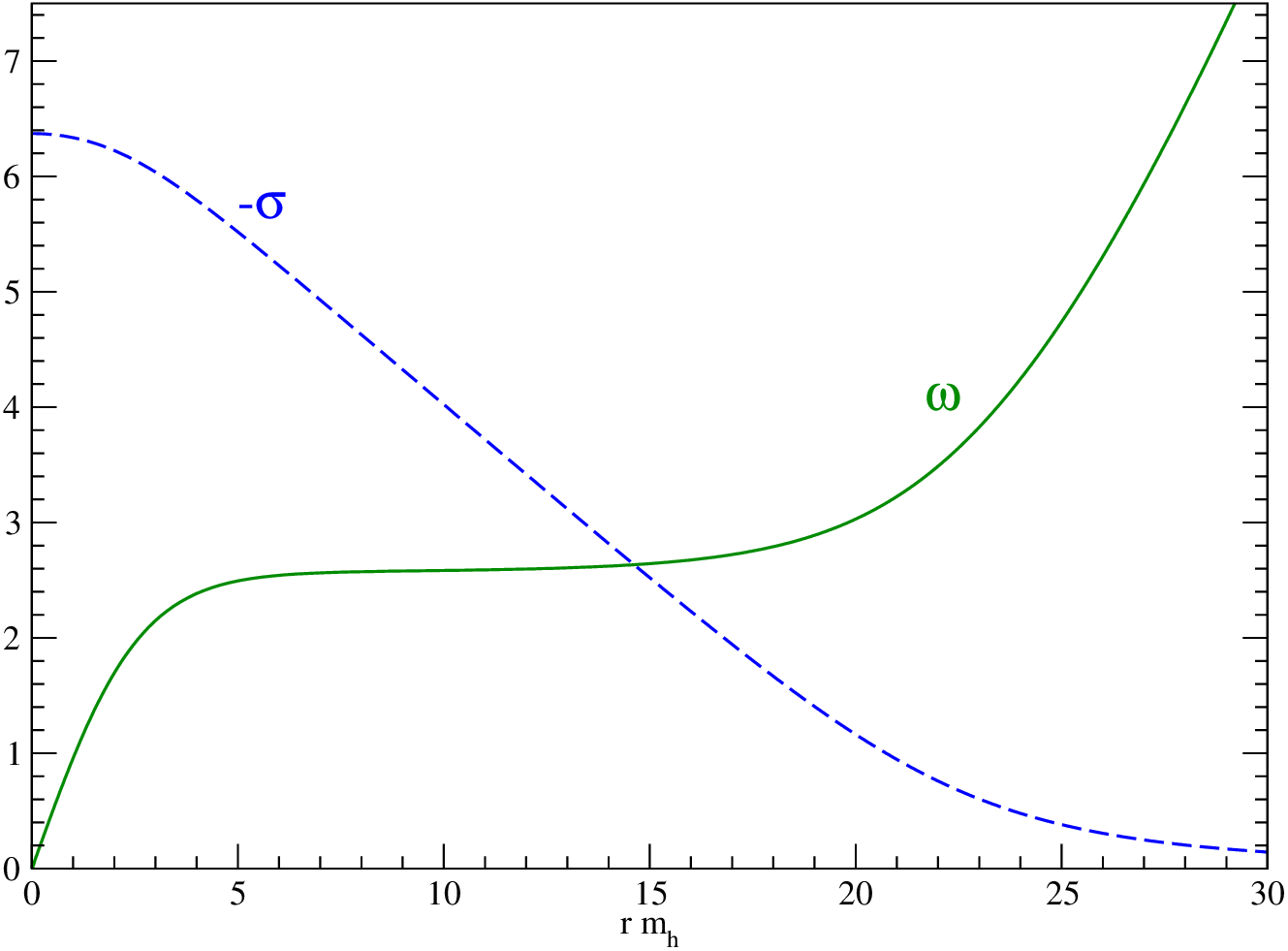}
\caption{$S^\ncomp$ hypersurface in the strongly curved region:
  $\omega(r)$ remains almost constant and the extra-dimensions becomes
  cylindrically shaped. The gravitational redshift in the core is
  given by $e^{\sigma(0)}$. The parameters are the same as in
  Fig.~\ref{fig:bgn3}.}
\label{fig:bgn3metric}
\end{center}
\end{figure}

In Fig.~\ref{fig:bgn3}, we have represented the field profiles
obtained for $\ncomp=3$, i.e.~in eight spacetime dimensions. The
dilaton condenses in the core as a scalar gravity field which
passively follows the stress energy distribution. The metric factor
$\sigma$ traces the gravitational redshift between the core and the
asymptotic spacetime and has been represented with $\omega(r)$ in
Fig.~\ref{fig:bgn3metric}. Finally, the Higgs and gauge fields are
typical of a topological defect configuration. By choosing $\epsilon =
0.25$, the gauge bosons are twice lighter than the Higgs boson and
condense within a larger extra-dimensional radius. The shift in the
condensation radius of the Higgs and gauge field produces the step
observed in the dilaton profile (see Fig.~\ref{fig:bgn3}).

\subsubsection{Dependence in the number of dimensions}

\begin{figure}
\begin{center}
\includegraphics[width=0.7\textwidth]{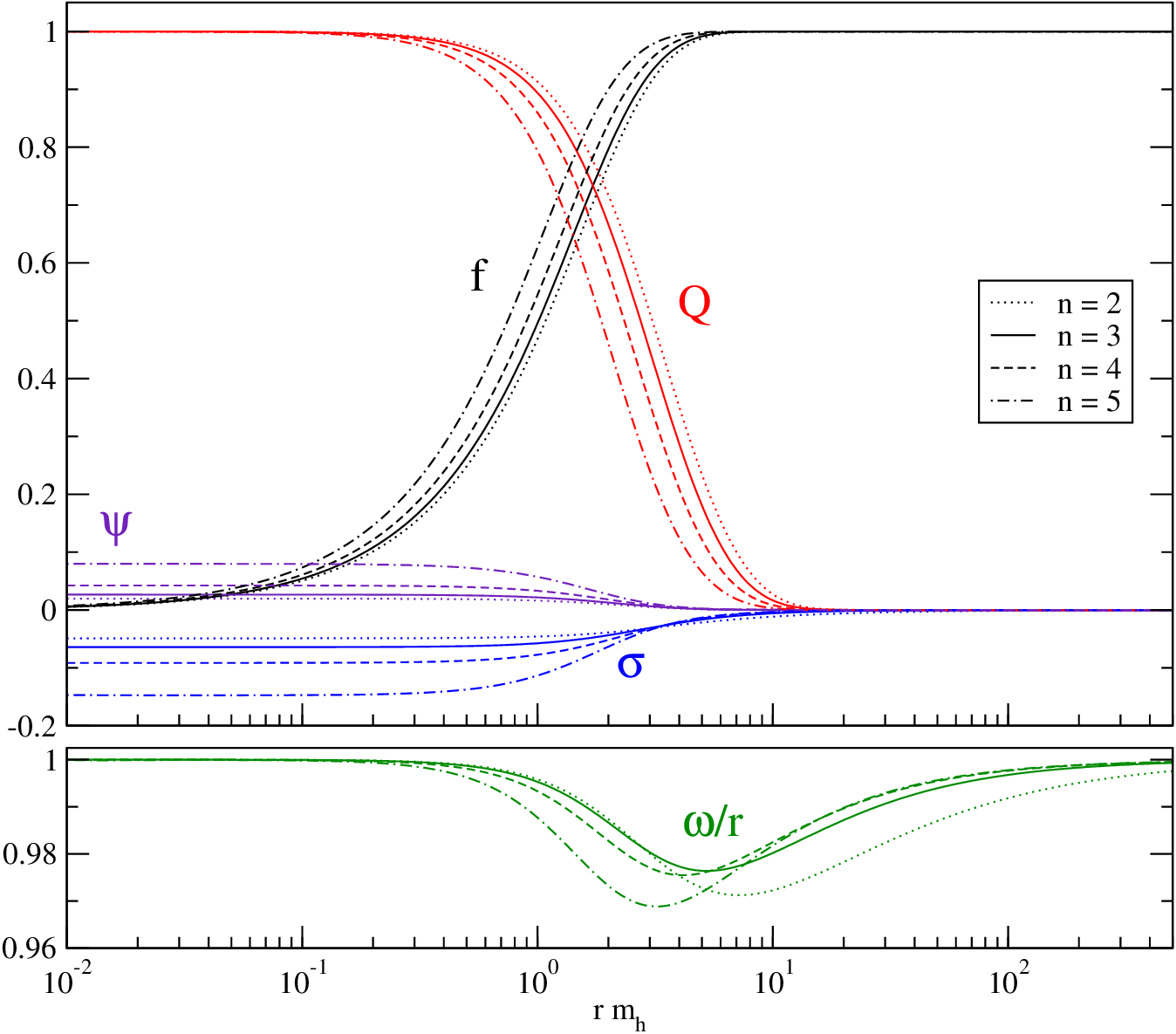}
\caption{Hypermonopole-forming field profiles in seven ($\ncomp=2$) to
  ten ($\ncomp=5$) dimensions. The coupling constants have been fixed
  to fiducial values $\alpha=0.1$, $\epsilon=0.2$ and $\beta =
  1$. Gravitational effects are enhanced by increasing $\ncomp$ with
  the exception of the metric coefficient $\omega(r)$. As can be
  checked in the lower panel, and in Eq.~(\ref{eq:enerperp}),
  deviations from flat space are first reduced by increasing $\ncomp$
  from $2$ to $3$, and then amplified for $\ncomp > 3$.}
\label{fig:bgn2to5}
\end{center}
\end{figure}

As discussed in the beginning of this section, the hypermonopole-forming fields
exhibit a similar behaviour for all values of $\ncomp$. In fact, this
can be understood from the equations of motion (\ref{eq:munu}) to
(\ref{eq:Q}). The number of dimensions enters these equations through
$\ncomp$ at two levels.

First, it changes the coupling between the metric coefficients
$\omadim$ and $\sigma$, as well as $\omadim$ and the other fields. For
a given $\omadim(r)$ profile, one expects all the fields to be more
sensitive to it when $\ncomp$ becomes large. Let us emphasise that
this effect is purely geometrical since it persists when $\alpha
\rightarrow 0$. In Fig.~\ref{fig:bgn2to5}, we have represented the
hypermonopole solutions obtained for an assumed generic set of
parameters $\alpha=0.1$, $\epsilon=0.2$, $\beta=1$ and in various
dimensions ranging from seven ($\ncomp=2$) to ten ($\ncomp=5$). The upper panel of this figure illustrates the
above-mentioned effect. For larger values of $\ncomp$, the
gravitational redshift $\sigma(0)$ increases while the dilaton $\psi$,
the gauge field $Q$ and the Higgs field $f$ condense in regions closer
to the hypermonopole core.

In the equations of motion, $\ncomp$ also affects the stress energy
tensor produced by the Higgs and gauge fields. As earlier mentioned,
the quantity $\calE_\perp$ in Eq.~(\ref{eq:enerperp}) has a dependence
in $(\ncomp-1)(\ncomp-4)$ which implies the same negative contribution
in seven ($\ncomp=2$) and eight-dimensions ($\ncomp=3$). However,
the kinetic terms are multiplied by a factor $(\ncomp-2)$ and vanish
in seven dimensions. The overall orthoradial energy has therefore a
non-trivial dependence in the number of dimension for $\ncomp \le 3$,
but should become monotonic for $\ncomp > 3$.  We can verify in the
lower panel of Fig.~\ref{fig:bgn2to5} that deviations from flat space
in the metric coefficient $\omega(r)$ are first reduced when $\ncomp$
goes from $2$ to $3$ before increasing again for $\ncomp > 3$.

In conclusion, we have found hypermonopole solutions for all tested
values of $\ncomp$. All the solutions exhibit the typical condensation
of Higgs and gauge fields encountered in topological defect
configurations. Since the dilaton also condenses, we do have an
effective varying Planck mass between the brane and the bulk. There is
also a strong gravitational redshift traced by the extra-dimensional
profile of $\sigma(r)$. Finally, at the matter field condensation
radius, the metric coefficient $\omega(r)$ remains almost constant and
the extra-dimensions become cylindrically shaped. In the next section,
we solve for the propagation of spin two fluctuations inside such
a background and show that gravitons becomes resonant on the length
scales at which the extra-dimensions are strongly curved.

\section{Resonant gravitons}
\label{sec:spintwo}

Restricting our attention to transverse and traceless four-dimensional
tensor fluctuations $h_{\mu\nu}$, the metric assumes the form given in
Eq.~(\ref{eq:metric}) with the replacement $\eta_{\mu\nu} \rightarrow
\eta_{\mu\nu} + h_{\mu \nu}$. Their linearised equation of motion can
be obtained by expanding the action in Eq.~(\ref{eq:action}) at second
order and has already been derived in
Ref.~\cite{Ringeval:2004ju}. Defining the conformal radial distance
$z(r)$ and the rescaled tensor fluctuations $\xi_{\mu\nu}$ by
\begin{equation}
z \equiv \mh \int_0^r e^{-\sigma(u)/2} \ud u, \qquad \xi_{\mu \nu}
\equiv e^{3 \sigma/4 + \psi/2} \omadim^{\ncomp/2} h_{\mu \nu},
\end{equation}
the equation of motion for the gravitons reads~\cite{Ringeval:2004ju}
\begin{equation}
\label{eq:mode}
-\xi''_{\mu\nu} + \left(W^2 + W' - \dfrac{e^\sigma}{\omadim^2}
  \Lapbel - \Boxadim \right) \xi_{\mu \nu} = 0,
\end{equation}
where a prime stands for derivative with respect to $z$. The quantity
$W(z)$ is a superpotential given by
\begin{equation}
W(z) \equiv \dfrac{3}{4} \sigma' + \dfrac{\ncomp}{2}
\dfrac{\omadim'}{\omadim} + \dfrac{1}{2} \psi',
\end{equation}
while $\Lapbel$ is the Laplace-Beltrami operator on the
$\ncomp$-sphere $S^\ncomp$
\begin{equation}
  \Lapbel \equiv \sum_{i=1}^{\ncomp}
  \dfrac{1}{\harm_i} \left( \dfrac{\partial^2}{\partial \theta_i^2} +
      \dfrac{\ncomp - i}{\tan \theta_i} \dfrac{\partial}{\partial
        \theta_i} \right),
\end{equation}
and $\Boxadim$ is the four-dimensional d'Alembertian
\begin{equation}
\Boxadim \equiv \dfrac{1}{\mh^2} \eta_{\mu \nu} \partial_\mu \partial_\nu\,.
\end{equation}
In order to solve Eq.~(\ref{eq:mode}), it is convenient to perform a
four-dimensional Fourier transform and a decomposition over the
hyperspherical harmonics $\Ylm{\ell}{m}$ such that the mode functions
$\umode{M}{\ell}(z)$ satisfy
\begin{equation}
\label{eq:umode}
- \umode{M}{\ell}'' + \left[W^2 + W' + \ell\left(\ell + \ncomp -1
  \right) \dfrac{e^\sigma}{\omadim^2} - M^2 \right] \umode{M}{\ell} = 0.
\end{equation}
Here $M^2$ and $-\ell(\ell+\ncomp-1)$ are the respective eigenvalues of
the d'Alembertian and Laplace--Beltrami operator (in Higgs mass
unit). This equation assumes the form of a Schr\"odinger equation of a
supersymmetric quantum mechanical system~\cite{Cooper:1994eh}. The
central potentials associated with fermionic- and bosonic-like
excitations are given by
\begin{equation}
V_2(z) = W^2 + W', \qquad V_1(z) = W^2 - W'.
\end{equation}
Omitting the tensor indices, the ground state $\xi_0$ of
Eq.~(\ref{eq:umode}) is the solution obtained for $M^2=\ell=0$ and
satisfies
\begin{equation}
\label{eq:susyground}
\left(\dfrac{\ud}{\ud z} + W\right) \left(-\dfrac{\ud}{\ud z } + W
\right)\xi_0 = 0,
\end{equation}
i.e.
\begin{equation}
\xi_0 \propto \omadim^{\ncomp/2} e^{3\sigma/4 + \psi/2}.
\end{equation}
This zero mode is not normalisable asymptotically. The ground state of
the superpartner potential $V_1$ is similarly obtained by swapping
both terms in Eq.~(\ref{eq:susyground}) and is given by $\tilde{\xi}_0
\propto 1/\xi_0$. This time, it is not normalisable in the
hypermonopole core for $z \rightarrow 0$. As a result, there are no
massless gravitons trapped on the brane and ``supersymmetry'' is broken
by the solutions we are interested in (the spectrum associated with
$V_2$ and $V_1$ do not match). Notice that the supersymmetric
properties of Eq.~(\ref{eq:umode}) ensures that the spectrum is
positive and no tachyonic propagation modes can be present.

In order to solve Eq.~(\ref{eq:mode}) in general, we assume that the
mode functions $\umode{M}{\ell}$ are normalised such that
\begin{equation}
\label{eq:umodenorm}
  \int_0^\infty \umode{M}{\ell}^*(z_1)
  \umode{M}{\ell}^{\phantom{*}}(z_2) \ud M 
  = \delta(z_1-z_2).
\end{equation}
It is now straightforward to check that the Green function for
$\xi_{\mu\nu}$ reads
\begin{eqnarray}
\label{eq:green}
\green{X_1;X_2} & = & - \int \dfrac{\ud^4 p}{(2 \pi)^4} e^{i p_\mu
    (x_1^\mu-x_2^\mu)} \nonumber \\ & \times & \sum_{\ell,\vect{m}}
  \Ylm{\ell}{m}(\vect{\theta_1}) \Ylm{\ell}{m}^*(\vect{\theta_2}) \int
  \dfrac{\umode{M}{\ell}^{\phantom{*}}(z_1)
    \umode{M}{\ell}^*(z_2)}{M^2 + \vec{p}^2 - (p^0 + i
    \varepsilon)^2}\ud M\,.
\end{eqnarray}
Capital letters have been used for $\Ntot$-dimensional coordinates,
bold characters for the $\ncomp$-dimensional vectors lying on
$S^\ncomp$, and arrows for the usual three-dimensional vectors on the
brane. Let us mention that we will not need to specify an explicit
expression for the $\Ylm{\ell}{m}$ functions and solely assume
they form an orthonormal basis such that
\begin{eqnarray}
\label{eq:orthoYlm}
  \int \left|\Ylm{\ell}{m}(\vect{\theta})\right|^2
  \calJ_\ncomp(\vect{\theta}) \, \ud \theta_1 \ud\theta_2 \dots
  \ud \theta_\ncomp = 1, \nonumber \\
  \sum_{\ell,\vect{m}} \Ylm{\ell}{m}(\vect{\theta_1})
  \Ylm{\ell}{m}^*(\vect{\theta_2}) = [\calJ_\ncomp(\vect{\theta_1})]^{-1} \,
  \delta^\ncomp(\vect{\theta_1} - \vect{\theta_2}),
\end{eqnarray}
where $\calJ_\ncomp(\vect{\theta}) \ud\theta_1 \cdots \ud\theta_n$ is the infinitesimal surface element on the unit
$\ncomp$-sphere so that
\begin{equation}
\label{eq:unitsurf}
  \calJ_\ncomp = \sin^{\ncomp-1}(\theta_1) \sin^{\ncomp-2}(\theta_2)
  \dots \sin(\theta_{\ncomp-1}).
\end{equation}

{}From the Green function, we can derive $h_{\mu\nu}$ for any additional
stress energy tensor on the brane. Considering an additional
transverse and traceless four-dimensional source $s_{\mu \nu}(x)$
inducing a $\Ntot$-dimensional linear stress-tensor perturbation of
the form (in Higgs mass unit)
\begin{equation}
\label{eq:tensorpert}
\delta T_{\mu \nu}(X) = \lim_{z\rightarrow 0}
\dfrac{\delta(z)}{z^\ncomp} [\calJ_\ncomp(\vect{\theta})]^{-1}
\delta^\ncomp(\vect{\theta}) s_{\mu\nu}(x),
\end{equation}
the tensor fluctuations at $X=X_1$ are given by Eq.~(\ref{eq:tensor})
and read
\begin{eqnarray}
\label{eq:greensol}
h_{\mu\nu}(X_1) & = & - \dfrac{2 \kappadim^2
    }{\omadim^{\ncomp/2}(z_1)} e^{-3 \sigma(z_1)/4}
  e^{-\psi(z_1)/2} \nonumber \\ &\times &\int \ud^\Ntot X_2  \green{X_1;X_2}
  \omadim^{\ncomp/2}(z_2) e^{3 \sigma(z_2)/4} e^{-\psi(z_2)/2} \delta
  T_{\mu\nu}(X_2),
\end{eqnarray}
with $\kappadim^2 \equiv \mh^{\ncomp+3} \kappa^2$.  The only unknowns
are the mode functions $\umode{M}{\ell}(z)$ entering the definition of
the Green function in Eq.~(\ref{eq:green}), and solution of
Eq.~(\ref{eq:umode}). It is instructive to solve them assuming
no-dilaton and flat space-time, i.e.~without the presence of the
hypermonopole.

\subsection{Flat spacetime}

Assuming $\psi=\sigma=0$ as well as $\omadim = z$ along the
extra-dimensions, Eq.~(\ref{eq:umode}) is a Bessel equations whose
regular solutions in the origin read~\cite{Gradshteyn:1965aa}
\begin{equation}
\label{eq:uflat}
\umodeflat{M}{\ell} = \sqrt{M z} \, \besselJ{\nu}(Mz),
\end{equation}
with
\begin{equation}
\nu \equiv \dfrac{\ncomp + 2 \ell - 1}{2}\,.
\end{equation}
The orthonormalisation properties of the Bessel
functions~\cite{Abramovitz:1970aa} automatically ensures that
Eq.~(\ref{eq:umodenorm}) is satisfied. Plugging Eq.~(\ref{eq:uflat})
into Eq.~(\ref{eq:greensol}) and looking for solutions sourced by
perturbations of the form~(\ref{eq:tensorpert}) gives
\begin{eqnarray}
h_{\mu\nu}^\flat(X_1) & = & \dfrac{2\kappadim^2 z_1^{(1-\ncomp)/2}}{2^\nu
    \Gamma(\nu + 1)} \sum_{\ell,\vect{m}}
  \Ylm{\ell}{m}(\vect{\theta_1})
  \Ylm{\ell}{m}(\vect{0})\left(\lim_{z_2 \rightarrow 0} z_2^{\ell}
  \right)   \nonumber \\ &\times&
    \int \ud^4 x_2\int \ud M M^{\nu+1} \besselJ{\nu}(M z_1) \Dret{M}{x_1-x_2} s_{\mu\nu}(x_2)\,.
\end{eqnarray}
The function $\Dret{M}{x}$ is the four-dimensional retarded propagator
defined by
\begin{eqnarray}
\Dret{M}{x}  & \equiv & \int \dfrac{\ud^4p}{(2
    \pi)^4} \dfrac{e^{i p_\mu x^\mu}}{M^2 + (\vec{p})^2 -
    (p^0 + i \varepsilon)^2} \nonumber \\
& = & \dfrac{\heavy{x^0}}{2\pi}\left[\delta(s^2) - \dfrac{M}{2s}
  \heavy{s^2} \besselJ{1}(M s)\right],
\end{eqnarray}
with $s^2\equiv (x^0)^2 - (\vec{x})^2$. The term in $z_2^{\ell}$
shows that only the hyperspherical harmonics with zero eigenvalues
$\ell=0$ contribute to the interactions sourced on the brane ($z_2=
0$). For static sources, performing the previous integrations and
evaluating the solution also on the brane ($z_1=0$) yields
\begin{equation}
\label{eq:hmunuflat}
\fl  h_{\mu\nu}^\flat(\vec{x}_1) = \dfrac{2 \kappadim^2}{4 \pi \mh^2}
  \dfrac{\left| \Ylm{0}{0}(\vect{0}) \right|^2}{2^{\ncomp-1}
    \left[\Gamma\negthinspace\left(\dfrac{\ncomp+1}{2}\right)
    \right]^2 } \int \ud^3
  \vec{x}_2 s_{\mu \nu}(\vec{x}_2) \times \int \ud M M^\ncomp
  \dfrac{e^{-M |\Delta \vec{x}|}}{|\Delta \vec{x}|}\,,
\end{equation}
where $\Delta \vec{x} \equiv \vec{x}_1 - \vec{x}_2$.  The last term in
the previous equation is the Laplace transform of $M^n$ which is
$\Gamma(\ncomp+1)/|\Delta \vec{x}|^{\ncomp+1}$. {}From
Eq.~(\ref{eq:orthoYlm}), one has
\begin{equation}
  \left|\Ylm{0}{0}(\vect{0}) \right|^2 = \left[\int
    \calJ_\ncomp(\vect{\theta}) \ud^\ncomp \vect{\theta}\right]^{-1} = \dfrac{1}{\calS^\ncomp}\,,
\end{equation}
where $\calS^\ncomp$ is the hypersurface of the unit
$\ncomp$-sphere:
\begin{equation}
  \calS^\ncomp = \dfrac{2
    \pi^{(\ncomp+1)/2}}{\Gamma\negthinspace\left(\dfrac{\ncomp+1}{2} \right)}\,.
\end{equation}
After having restored the dimensions, Eq.~(\ref{eq:hmunuflat})
simplifies into
\begin{equation}
  h_{\mu\nu}^\flat(\vec{x}_1) = \dfrac{2 \kappa^2}{(\ncomp+2)
    \calS^{\ncomp+3}} \int \ud^3 \vec{x}_2\dfrac{s_{\mu
      \nu}(\vec{x}_2)}{\left|\vec{x}_1 - \vec{x}_2\right|^{\ncomp+2}}\,,
\end{equation}
which is the standard linearised solution of the Einstein equations in
$\Ntot = 5 + \ncomp$ spacetime dimensions. The $N$-dimensional Newton
constant also matches with the standard value
\begin{equation}
\label{eq:Gnewton}
G_N = \dfrac{\kappa^2}{(\ncomp+3) \calS^{\ncomp+3}}\,.
\end{equation}

\subsection{Inside the hypermonopole}

Inside the hypermonopole, the tensor fluctuations can be derived in a
similar way. One should first keep the factors involving $\omadim$,
$\sigma$ and $\psi$. In fact, as can be seen from
Eq.~(\ref{eq:greensol}), by taking both the source and the observer on
the brane, all factors involving $\sigma$ and $\omadim$ cancel, solely
the dilaton rescales the gravitational coupling constant by
$\exp[-\psi(0)]$. The mode function $\umode{M}{\ell}$ are no longer
the same but for both the source and the observer on the brane, only
their value in $z=0$ enters the calculation. Furthermore, since the
only hyperspherical harmonic which is non-zero at $\bm{\theta}=0$ is
$f_0^{\bm{0}}$, only the $l=0$ modes contribute to the tensors
fluctuations. In fact, by defining the spectral density
\begin{equation}
\rho(M) \equiv
\dfrac{\left|\umode{M}{0}^{\phantom{\flat}}(0)\right|^2}{ \left| \umode{M}{0}^\flat(0) \right|^2}\,,
\end{equation}
one arrives at
\begin{equation}
\label{eq:hmunu}
  h_{\mu\nu}(\vec{x}_1) = \dfrac{2 \kappa^2 } {(\ncomp+2)
    \calS^{\ncomp+3}} \dfrac{e^{-\psi(0)}}{\Gamma(\ncomp+1)} 
  \int \ud^3 \vec{x}_2 \dfrac{ s_{\mu\nu}(\vec{x}_2)} {|\Delta
    \vec{x}|}\laplace{|\Delta \vec{x}|}{M^n \rho(M)} \,,
\end{equation}
where we have defined the Laplace transform
\begin{equation}
\laplace{x}{q(M)} \equiv \int_0^\infty e^{-M x} q(M)\, \ud M .
\end{equation}
A four-dimensional behaviour can be recovered if the Laplace transform
has a weak dependence in $|\Delta \vec{x}|$, which is precisely the
case when gravitons become resonant with a mass $\mg$. Taking as a toy
example 
\begin{equation}
\label{eq:rhotoy}
  \rho(M) = \varrho_0 + C \mg \delta(M-\mg),
\end{equation}
where $\varrho_0$ and $C$ are two constants, one gets
\begin{equation}
\label{eq:laptoy}
  \laplace{|\Delta \vec{x}|}{M^n \rho(M)} = \dfrac{\varrho_0\Gamma(\ncomp+1)}
  {|\Delta \vec{x}|^{\ncomp+1}} + C \mg^{\ncomp+1} e^{-\mg |\Delta \vec{x}|},
\end{equation}
whose second term dominates and is almost constant in the range
\begin{equation}
\label{eq:4Dbounds}
\left[\dfrac{\varrho_0\Gamma(\ncomp+1)}{C} \right]^{1/(\ncomp+1)} \ll  \mg|\Delta \vec{x}| \ll 1
\end{equation}
provided $\varrho_0$ is sufficiently small. Outside of this range, the inverse power term dominates and the tensor fluctuations are that of $N$-dimensional gravity.
The upper bound in Eq. (\ref{eq:4Dbounds}) is satisfied for graviton resonances which are light
enough, i.e. $\mg \rightarrow 0$, whereas the lower bound requires a
strong peaked resonance, i.e. a long lived graviton having $C \gg
\varrho_0$. In the following, we show that such a situation
generically occurs inside the hypermonopoles: gravitons become
strongly resonant due to the positive curvature of the
extra-dimensions.

\subsection{Graviton spectral density}

\begin{figure}
\begin{center}
\includegraphics[width=0.7\textwidth]{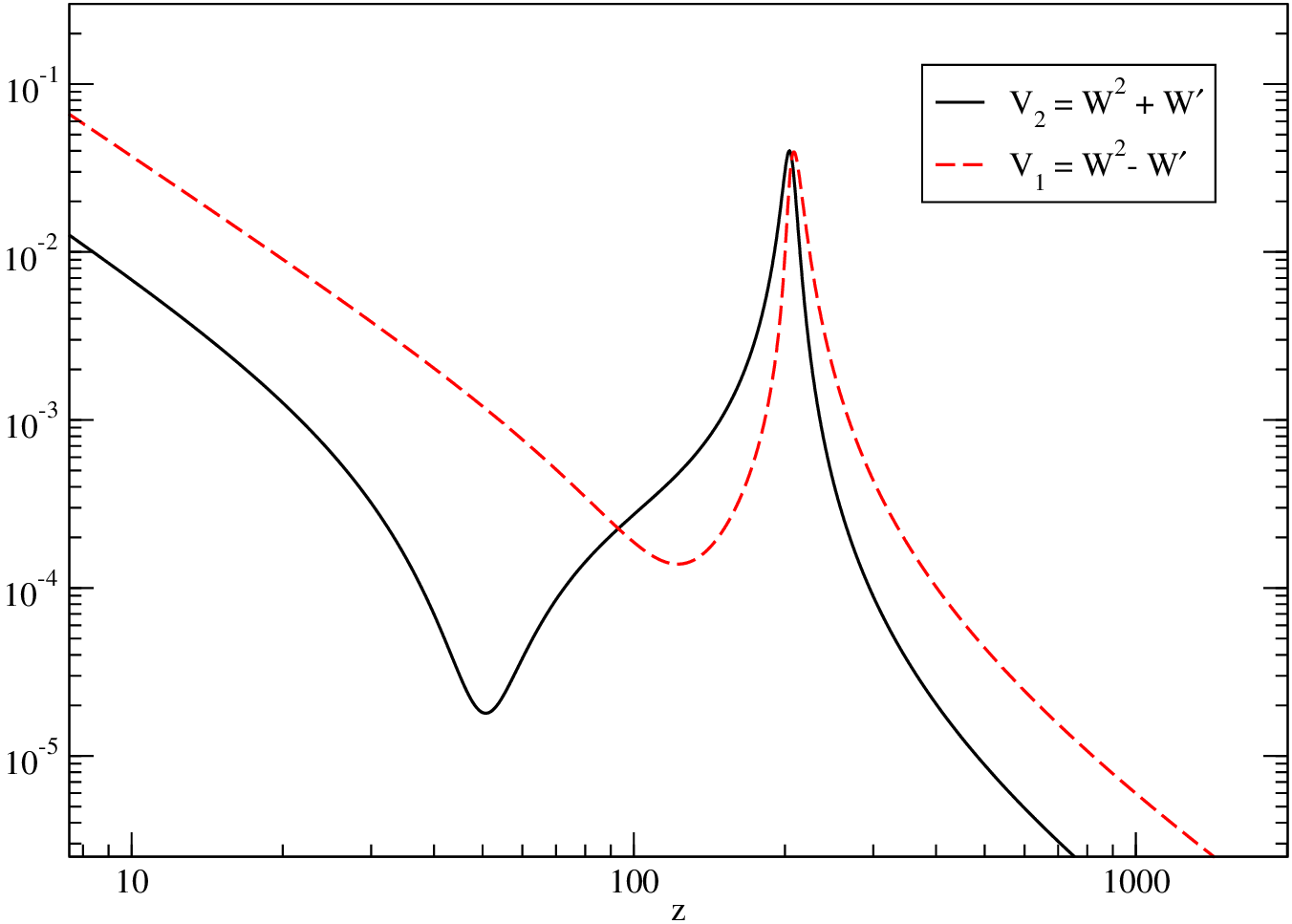}
\caption{Superpartner potentials $V_1$ and $V_2$ for the
  eight-dimensional hypermonopole of Sec.~\ref{sec:bg}. Although they
  do not support bound states, metastable massive gravitons are
  trapped in the confining nest around $z\simeq 200$. The peaks are
  associated with the regions of maximum curvature of
  Fig.~\ref{fig:bgn3} (same parameters).}
\label{fig:potentials}
\end{center}
\end{figure}

\begin{figure}
\begin{center}
\includegraphics[width=0.7\textwidth]{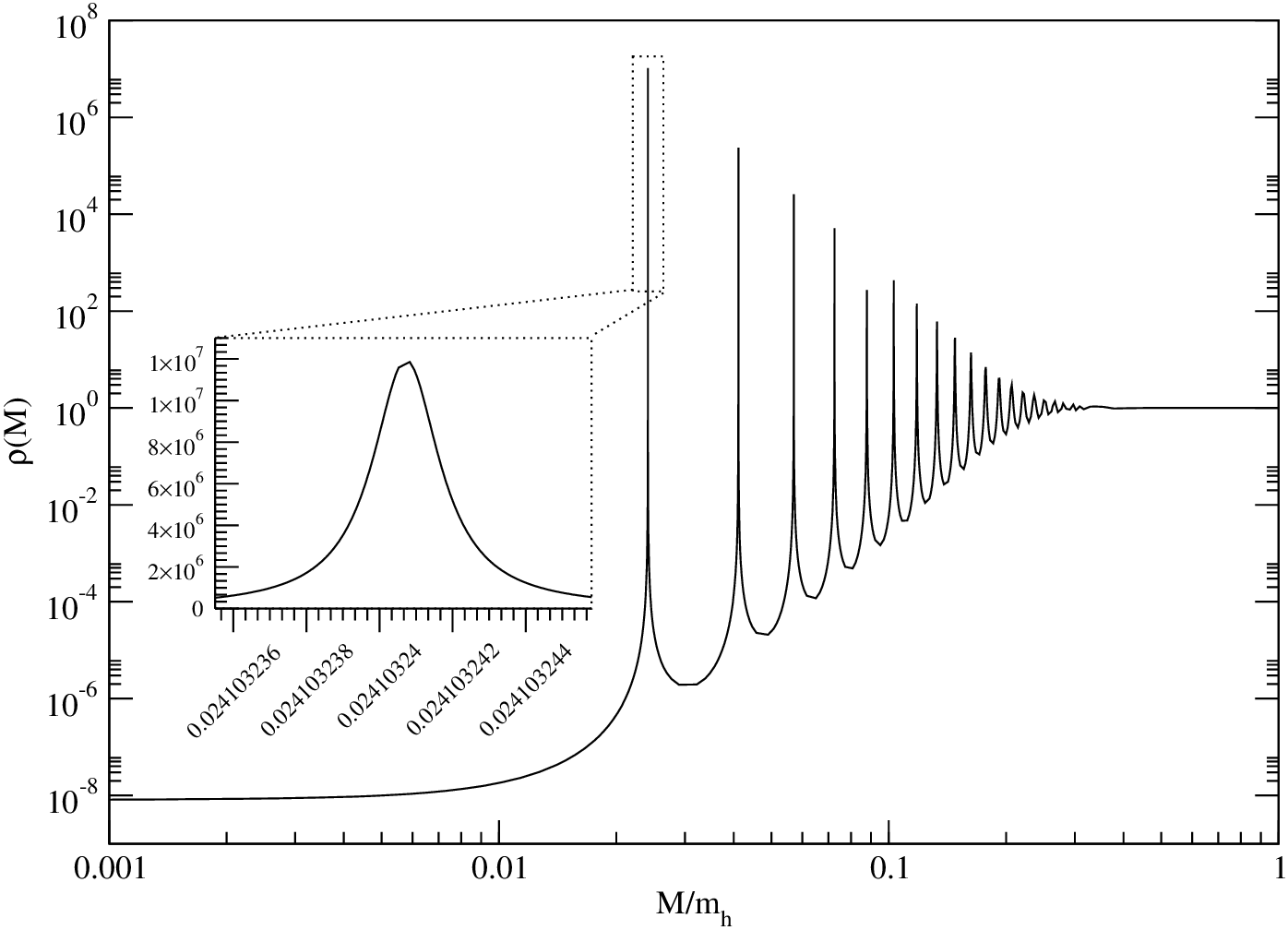}
\caption{Spectral density with respect to the graviton mass for the
  eight-dimensional hypermonopole. The resonances correspond to the
  massive metastable states associated with the potential $V_2(z)$
  plotted in Fig.~\ref{fig:potentials}. More than fifteen gravitons
  end up being trapped. The constant behaviour at small and large
  distances shows that gravity is $\Ntot$-dimensional in these
  regions, albeit with a different effective Newton constant due to
  the gravitational redshift.}
\label{fig:rho}
\end{center}
\end{figure}

Since the extra-dimensions are asymptotically flat, the mode functions
far from the hypermonopole core are of the form given by
Eq.~(\ref{eq:uflat}). Using these Bessel wave-form as asymptotic
initial conditions, we have numerically solved Eq.~(\ref{eq:umode}) in
the background fields of Fig.~\ref{fig:bgn3}. In
Fig.~\ref{fig:potentials}, we have represented the superpartner
potential $V_1$ and $V_2$ as a function of the conformal radial
distance $z$. Solving for the mode equation in $V_2(z)$ gives the
spectral density plotted in Fig.~\ref{fig:rho}. For this
configuration, we have observed more than fifteen trapped gravitons,
the lightest having a spectral density profile typical of a
Breit--Wigner distribution
\begin{equation}
\label{eq:bw}
  \rho(M) \underset{\mg}{\simeq} \dfrac{I} 
  {1 + \left(\dfrac{M - \mg}{\gammag} \right)^2}\,,
\end{equation}
whose best fit gives
\begin{eqnarray}
\label{eq:bestbw}
\mg &=& 2.4103240725 \times10^{-2} \mh, \nonumber \\
\gammag &=& 1.12\times 10^{-9} \mh \nonumber \\
I & =& 1.19 \times 10^7.
\end{eqnarray}
In order to properly resolve the width of these resonance, we have
implemented a recursive local adaptive mesh refinement coupled to the
more usual Runge--Kutta integration of Eq.~(\ref{eq:umode}). For such
very thin resonances, the toy model of Eq.~(\ref{eq:rhotoy}) is a good
approximation for $M\simeq\mg$, the constant $C$ being given by the
integral of Eq.~(\ref{eq:bw}). One finds
\begin{equation}
C = I\dfrac{\gammag}{2 \mg} \left(\pi + 2 \arctan\dfrac{\mg}{\gammag}
\right) \simeq \pi I \dfrac{\gammag}{\mg}\,,
\end{equation}
where the last expression is accurate only for $\gammag \ll \mg$. For
the best fit values of Eq.~(\ref{eq:bestbw}), one finds $C\simeq
1.7$. As can be seen in Fig.~(\ref{fig:rho}), for light masses
$\rho(M) = \varrho_0 = 8.2\times 10^{-9}$ such that the lower bound in
Eq.~(\ref{eq:4Dbounds}) is about $10^{-2}$. We therefore expect this
resonance to change the standard eight-dimensional gravity law on
distances covering not more that two-orders of magnitude around the
scale $1/\mg$, which is far to short to be interesting for cosmological
purpose.

However, we do not see any reasons preventing the existence of
cosmologically interesting solutions, i.e.~much lighter
gravitons. Indeed, increasing $\alpha$ appears to push up the
potential barrier in Fig.~\ref{fig:potentials} whereas reducing
$\epsilon$ increases the width of the potential barrier. Small values
of $\epsilon$ have the effect of delocalising the gauge field and this
ends up spreading its energy density over the extra-dimensions. Both
of these parameters could therefore be somehow adjusted to obtained
much lighter graviton resonances. As the numbers reported in
Eq.~(\ref{eq:bestbw}) suggest, the precise determination of lighter
resonances is made difficult due to numerical limitations, the machine
precision accuracy not covering more than $16$ orders of magnitude on
usual computers is already saturated by $\gammag/I$ in
Eq.~(\ref{eq:bestbw}).

\subsection{Deviations from Newton}

\subsubsection{Dimensional reduction}

In order to complete the discussion of the previous section, we have
computed the Laplace transform directly from the spectral density
found in Fig.~\ref{fig:rho}.
\begin{figure}
\begin{center}
\includegraphics[width=0.75\textwidth]{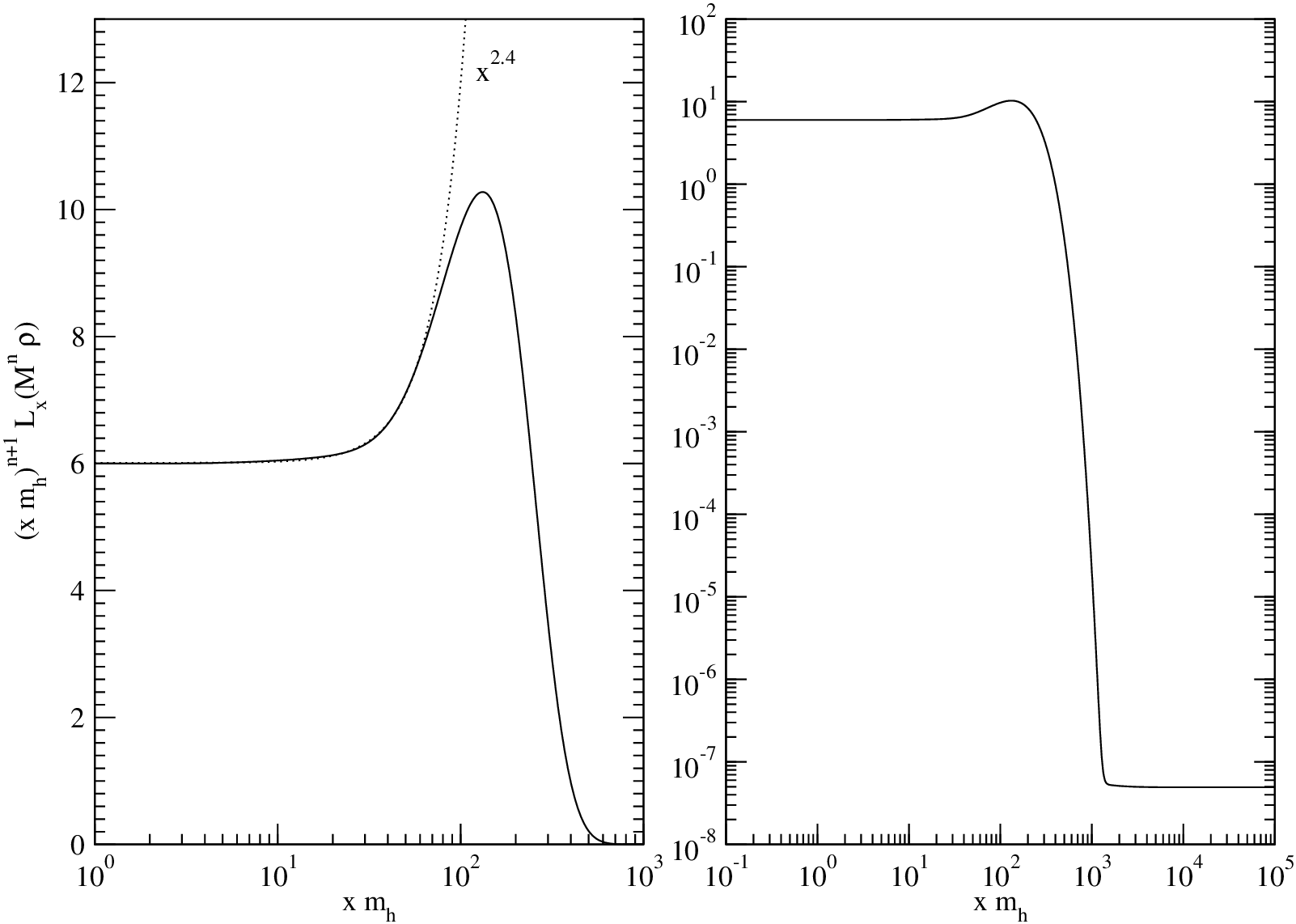}
\caption{Rescaled Laplace transform in linear (left) and logarithmic
  scale (right) associated with the spectral density of
  Fig.~\ref{fig:rho} and plotted as a function of the distance to the
  source. The Newton law is eight-dimensional when this quantity is
  constant as it occurs on small and large distances. The lightest
  graviton resonance of Eq.~(\ref{eq:bestbw}) reduces locally the
  dimensionality of the Newton force by a fractional power of exponent
  $2.4$ (best fit). The strong variation in amplitude from small to
  large scales comes from the hypermonopole induced gravitational
  redshift.}
\label{fig:newtn3}
\end{center}
\end{figure}
Fig.~\ref{fig:newtn3} shows the rescaled quantity
$|\vec{x}|^{\ncomp+1} \laplace{|\vec{x}|}{M^\ncomp \rho(M)}$ as a
function of the distance to the source. Gravity is $\Ntot$-dimensional
when this quantity is constant as it occurs at small and large
distances. The strong variation in amplitude, also visible in the
smooth change of the spectral density in Fig.~\ref{fig:rho}, comes
again from the hypermonopole gravitational redshift (see the $\sigma$
profile in Fig.~\ref{fig:bgn3}). The graviton resonance of
Eq.~(\ref{eq:bestbw}) is responsible of the peak located around $x
\simeq 1/\mg$. A best power fit of the potential nearby this region
shows that the Newton law is dimensionally reduced to
$1/x^{3.6}$. This is not yet a four-dimensional Newton law due
to the previously discussed numerical limitations to obtain a light
enough resonance.

\subsubsection{Effective gravitational coupling}

{}From Eqs.~(\ref{eq:Gnewton}), (\ref{eq:hmunu}) and (\ref{eq:laptoy}),
one can extract the effective Newton constant $G$ which would be
measured by a four-dimensional observer in the three expected
regimes. At very small distances,
\begin{equation}
  |\Delta\vec{x}| \mg \ll \left(\dfrac{\varrho_0 \Gamma(\ncomp+1)}{C}\right)^{1/(\ncomp+1)},
\end{equation}
the spectral density is constant $\rho(M)=1$ and thus gravity is seven
dimensional. The measured Newton constant is however reduced by the
dilaton condensation in the hypermonopole core:
\begin{equation}
G_0 = e^{-\psi(0)} G_\Ntot\,.
\end{equation}
In the intermediate regions, those verifying Eq.~(\ref{eq:4Dbounds}),
gravity is driven by the metastable lightest graviton and the Newton
law is four-dimensional with an effective Newton constant given by
\begin{equation}
\label{eq:Gerard}
G_4 = e^{-\psi(0)} \dfrac{G_\Ntot}{2}  \dfrac{n+3}{n+2} \dfrac{C}{\Gamma(\ncomp+1)} \mg^{\ncomp+1}.
\end{equation}
This equation makes clear that a light graviton, required for the
dimensional reduction of the Newton law, will necessarily induce a
small effective Newton constant thereby addressing the mass hierarchy
problem. In terms of the reduced Planck masses, Eq.~(\ref{eq:Gerard})
can be recast into
\begin{equation}
\label{eq:Max}
\Mp^2 = 2 e^{\psi(0)} \dfrac{n+2}{n+3} \dfrac{\Gamma(\ncomp+1)}{C} \dfrac{M^{\ncomp+3}}{\mg^{\ncomp+1}}\,,
\end{equation}
where $\Mp^2 = 1/G_4$ and $M^{\Ntot-2}=1/G_\Ntot$.

Finally, on the largest length scales,
\begin{equation}
|\Delta\vec{x}| \mg \gg 1,
\end{equation}
gravity becomes again $\Ntot$-dimensional but with a much weaker
Newton constant since now $\rho(M \rightarrow 0) = \varrho_0 \ll
1$. The measured Newton constant is now given by
\begin{equation}
  G_\infty = \varrho_0 e^{-\psi(0)}  G_\Ntot = \varrho_0 G_0 \ll G_0.
\end{equation}
As a numerical application, we can determine the order of magnitude of
the $\Ntot$-dimensional Planck mass such that the graviton mass is of
the same order than the cosmological constant energy scale,
i.e. $\mg\simeq 10^{-11}\GeV$. {}From Eq.~(\ref{eq:Max}), one gets
\begin{equation}
  M \simeq \left[\dfrac{C
      e^{-\psi(0)}}{\Gamma(\ncomp+1)}\right]^{1/(\ncomp+3)}
  \times 10^{\frac{25-11\ncomp}{\ncomp+3}}\GeV,
\end{equation}
which is down to the $10\, \GeV$ scale already for $\ncomp=2$. Notice
that the lowest scale at which $\Ntot$-dimensional gravity shows up is
fixed by the value of $\varrho_0$ and not by $\mg$. This quantity
coming only from the gravitational redshift $\sigma(0)$, it can
actually be made arbitrarily small for order one coupling constants,
i.e.~for Higgs vacuum expectation values also around the $10\,\GeV$
scale. In fact, as suggested by the previous equation, the mass
hierarchy mechanism advocated here is so efficient that it has a
natural preference for very small $\Ntot$-dimensional Planck masses.

\section{Conclusion}
 
In this paper we have shown that metastable massive gravitons
generically exist in the four-dimensional core of any self-gravitating
hypermonopoles formed by the breakdown of an $SO(\nextra)$ symmetry in
$(\nextra+4)$ dimensions, provided $\nextra \ge 3$.

Since the extra-dimensional spacetime is of infinite volume and
asymptotically flat, these resonances induce a DGP-like gravity
confinement mechanism in the core. For light enough resonances,
gravity is $\Ntot$-dimensional at small and large distances, but can
be four-dimensional on some intermediate range. The numerical
determination of such a light and long-lived resonance may be however
a non-trivial problem due to finite numerical accuracy. Moreover, we
have shown that, in this regime, the effective four-dimensional Planck
mass is proportional to an inverse power of the graviton mass; this
one being extremely light, the mass hierarchy problem ends up being
naturally addressed in our setup. The strong decay of the gravity law
at large distances, coming from both the higher-dimensionality and the
strong gravitational redshift, might be of interest to explain the
current cosmic acceleration.

Finally, we would like to emphasize that these models still remain
unexplored on various aspects which may compromise, or not, their
viability. Here, we have only solved the propagation of spin two
fluctuations which decouple from the background fields. The model has
however vector and scalar modes which may propagate and might also be
confined in the core. Solving for their propagation is a challenging
problem since they will be necessarily coupled to all of
hypermonopole-forming vector and scalar fields. We leave the second
order perturbation of Eq.~(\ref{eq:action}) in the scalar and vector
modes for a future work.

\begin{acknowledgments}
  This work is supported by the Belgian Federal Office for Science,
  Technical and Cultural Affairs, under the Inter-university
  Attraction Pole Grant No. P6/11.
\end{acknowledgments}

\section{Appendix}

In this appendix, we present the Einstein tensor $G_{AB}$ and the
stress tensor $T_{AB}$. Inserted into Eq.~(\ref{eq:tensor}), we then
arrive at the equations of motion for the gravity sector
(\ref{eq:munu}), (\ref{eq:rr}) and (\ref{eq:theta}).  For the metric
given in Eq.~(\ref{eq:metric}), the Einstein tensor is given by
\begin{eqnarray}
  G_{\mu\nu} & = & \mh^2 g_{\mu\nu}\left[
    \frac{3}{2}\ddot{\s}+\frac{3}{2}\dot{\s}^2+\ncomp\frac{\ddot{\bo}}{\bo}
    +\frac{\ncomp(\ncomp-1)}{2}\left(\frac{\dot{\bo}}{\bo}\right)^2  \frac{3\ncomp}{2}\frac{\dot{\s}\dot{\bo}}{\bo}-\frac{\ncomp(\ncomp-1)}{2\bo^2}\right], \nonumber \\
  G_{rr}& = & \mh^2\left[\frac{3}{2}(\dot{\s})^2 +
    \frac{\ncomp(\ncomp-1)}{2}\left(\frac{\dot{\bo}}{\bo}\right)^2+2\ncomp\frac{\dot{\s}\dot{\bo}}{\bo}
    -\frac{\ncomp(\ncomp-1)}{2\bo^2}\right], \nonumber \\
  G_{\theta_i\theta_j}& = & \mh^2 g_{\theta_i\theta_j}\left[2\ddot{\s}
    +\frac{5}{2} (\dot{\s})^2 +
    (\ncomp-1)\frac{\ddot{\bo}}{\bo} \right. \nonumber \\ &+& \left. 2(\ncomp-1)
    \frac{\dot{\s}\dot{\bo}}{\bo}
    \frac{(\ncomp-1)(\ncomp-2)}{2}\left(\frac{\dot{\bo}}{\bo}\right)^2-\frac{(\ncomp-1)(\ncomp-2)}{2\bo^2}\right].
\end{eqnarray}

With the Higgs and gauge fields specified by Eqs.~(\ref{eq:anshiggs}),
(\ref{eq:fdefinition}) and (\ref{eq:ansgauge}), the stress tensor
(\ref{eq:tmunugen}) becomes
\begin{eqnarray}
  T_{\mu\nu} & = & \frac{\vev^2\mh^2}{2} g_{\mu\nu}
  \left[-\dot{f}^2-\ncomp\left(\frac{f Q}{\bo}\right)^2
    -\frac{\ncomp}{ \epsilon \bo^2} 
    \dot{Q}^2  - \frac{\ncomp(\ncomp-1)}{2 \epsilon \bo^4}(1-Q^2)^2
    \nonumber \right. \\  &-& \left. \frac{1}{4}(f^2-1)^2 \right],
  \nonumber \\
  T_{rr} &=&  \frac{\vev^2\mh^2}{2}\left[\dot{f}^2
    +\frac{\ncomp}{\epsilon \bo^2} \dot{Q}^2-\ncomp\left(\frac{f
        Q}{\bo}\right)^2
    - \frac{\ncomp(\ncomp-1)}{2 \epsilon \bo^4}(1-Q^2)^2 \nonumber
  \right. \\ &-& \left.  \frac{1}{4}(f^2-1)^2\right], \nonumber \\
  T_{\theta_i \theta_j} &= & \frac{\vev^2\mh^2}{2} g_{\theta_i
    \theta_j}\left[-\dot{f}^2-(\ncomp-2)\left(\frac{f Q}{\bo}\right)^2
    \nonumber \right. \\ & -& \left. \frac{\ncomp-2}{\epsilon\bo^2}
    \dot{Q}^2 - \frac{(\ncomp-1)(\ncomp-4)}{2\epsilon
      \bo^4}(1-Q^2)^2-\frac{1}{4}(f^2-1)^2\right].
\end{eqnarray}

\section*{References}
\bibliography{bibpole}

\end{document}